# Comment on “Resolution of the Abraham-Minkowski Dilemma”

Changbiao Wang*
ShangGang Group, 70 Huntington Road, Apartment 11, New Haven, CT 06512, USA

In a recent Letter by Barnett [S. M. Barnett, Phys. Rev. Lett. 104, 070401 (2010)], a total-momentum model is proposed for resolution of the Abraham-Minkowski dilemma. In this model, Abraham's and Minkowski's momentums are, respectively, a component of the same total momentum, with the former being the kinetic momentum and the latter the canonical momentum. In this Comment, I would like to indicate that this physical model is not consistent with global momentum-energy conservation law in the principle-of-relativity frame.

In a recent Letter by Barnett [1], a total-momentum model is proposed for resolution of the Abraham-Minkowski dilemma. In this model, Abraham’s and Minkowski’s momentums are, respectively, a component of the same total momentum, as shown in his Eq. (7) given by $\mathbf{p}_{\text{kin}}^{\text{med}}+\mathbf{p}_{\text{Abr}}=\mathbf{p}_{\text{can}}^{\text{med}}+\mathbf{p}_{\text{Min}}$ , with $\mathbf{p}_{\text{Abr}}$ being the Abraham’s (kinetic) momentum and $\mathbf{p}_{\text{Min}}$ the Minkowski’s (canonical) momentum. In this Comment, I would like to indicate that this physical model is not consistent with global momentum-energy conservation law in the principle-of-relativity frame. My arguments are given below.

According to the principle of relativity, the laws of physics are the same in all inertial frames of reference, and there is no preferred inertial frame for descriptions of physical phenomena. Thus Maxwell equations and global momentum-energy conservation law must be valid in all inertial frames, no matter whether the space is filled with dielectric materials partially or fully, and no matter whether the materials are moving or at rest. However when applying Barnett’s total-momentum model to analysis of medium Einstein-box thought experiment (also known as “Balazs thought experiment”) with the principle of relativity, one may find that the global momentum-energy conservation law is broken. The proof is given below.

(i) Suppose that *before* the photon enters the medium box, the photon *initially* is located far away from the medium box in *vacuum*. Thus initially the photon’s Abraham (= Minkowski) momentum and energy, $(\mathbf{p}_{\text{Abr}},E_{\text{Abr}}/c)_{before}$, constitute a Lorentz four-vector, as shown in Fig. 1(a).

(ii) The medium box is made up of *massive* particles, and its kinetic momentum and energy $(\mathbf{p}_{\text{kin}}^{\text{med}},E^{\text{med}}/c)$ constitute a Lorentz four-vector no matter *before* or *after* the photon enters the medium box.

(iii) From (i) and (ii), initially the total momentum and energy constitute a four-vector, namely

$$(\mathbf{p}_{\text{total}},E_{\text{total}}/c)_{before}=(\mathbf{p}_{\text{Abr}},E_{\text{Abr}}/c)_{before}+(\mathbf{p}_{\text{kin}}^{\text{med}},E^{\text{med}}/c)_{before} \quad (1)$$

is a four-vector.

* changbiao_wang@yahoo.com

(iv) After the photon enters the medium box, as shown in Fig. 1(b), $(\mathbf{p}_{\rm Abr}, E_{\rm Abr}/c)_{after}$ *cannot* be a four-vector according to the principle of relativity [2], but $(\mathbf{p}^{\rm med}_{\rm kin}, E^{\rm med}/c)_{after}$ is still a four-vector from (ii). Thus

$$(\mathbf{p}_{\rm total}, E_{\rm total}/c)_{after} = (\mathbf{p}_{\rm Abr}, E_{\rm Abr}/c)_{after} + (\mathbf{p}^{\rm med}_{\rm kin}, E^{\rm med}/c)_{after} \qquad (2)$$

*cannot be* a four-vector.

(v) From (iii) and (iv) we conclude that $(\mathbf{p}_{\rm total}, E_{\rm total}/c)_{after} = (\mathbf{p}_{\rm total}, E_{\rm total}/c)_{before}$ cannot hold in *all* inertial frames, namely the global momentum and energy conservation laws are broken in the principle-of-relativity frame.

From above analysis we can see that Barnett's total-momentum model is indeed not consistent with the principle of relativity and the global momentum-energy conservation law, which are all fundamental postulates in physics.

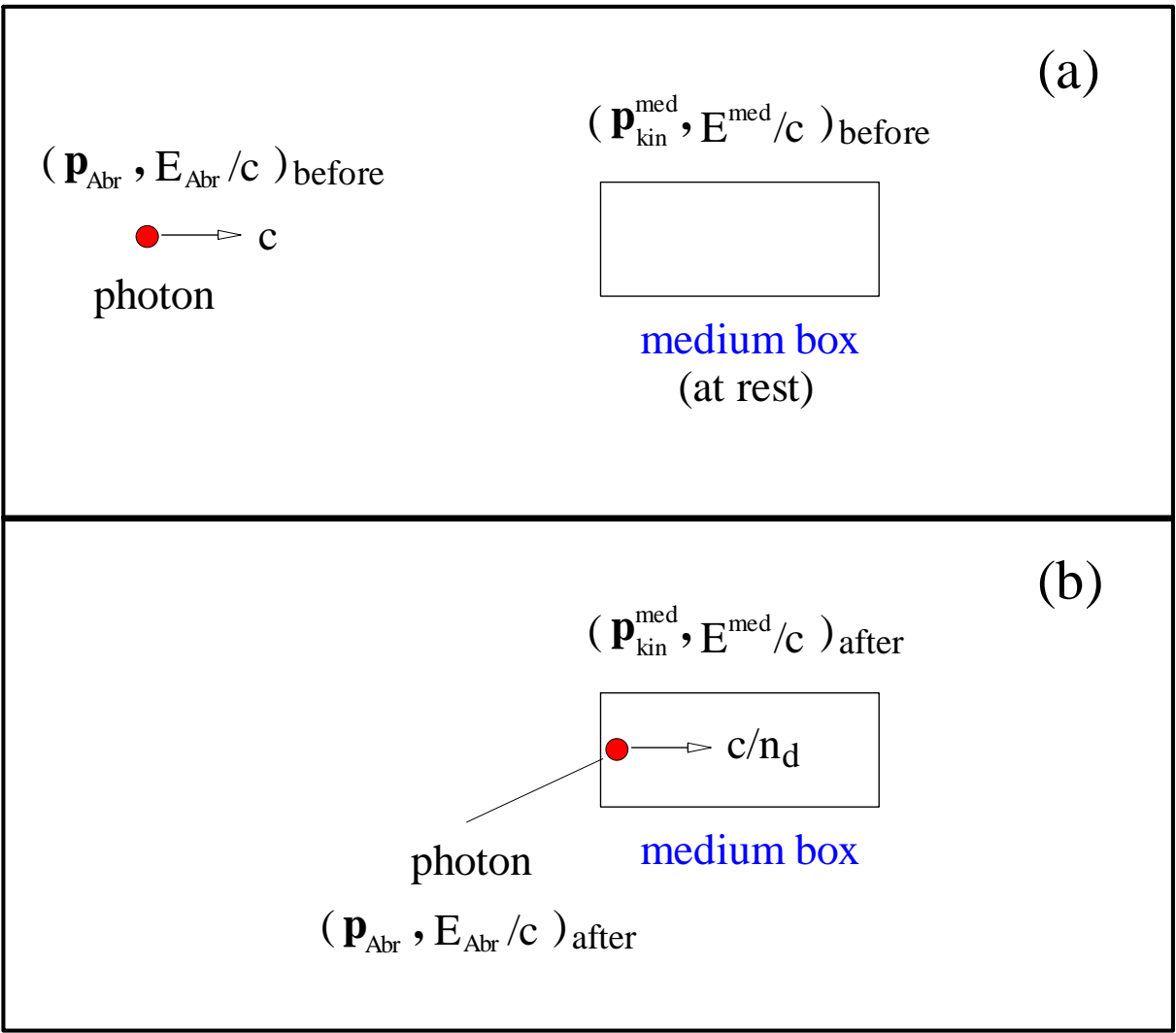

FIG. 1. Analysis of Einstein-box thought experiment (also known as "Balazs thought experiment") with the principle of relativity. (a) Initially the photon is located far away from the medium box in *vacuum*. The photon's Abraham (= Minkowski) momentum and energy $(\mathbf{p}_{\rm Abr}, E_{\rm Abr}/c)_{before}$ constitute a Lorentz four-vector. $(\mathbf{p}^{\rm med}_{\rm kin}, E^{\rm med}/c)_{before}$ is also a four-vector. (b) After the photon enters the medium box, the photon's Abraham momentum and energy $(\mathbf{p}_{\rm Abr}, E_{\rm Abr}/c)_{after}$ *cannot* be a four-vector, while $(\mathbf{p}^{\rm med}_{\rm kin}, E^{\rm med}/c)_{after}$ is still a four-vector. Note: $n_d$ is the refractive index of the medium box, and $c$ is the vacuum light speed.

There is a question one might be interested: How did Barnett get his total-momentum model flawed? The answer is given below.

In Barnett's model [1], the argument for supporting Abraham momentum is based on the analysis by "center-of-mass-energy" approach, where the global momentum-energy conservation law is *indeed* employed to obtain Abraham photon momentum and energy in the medium box in lab frame [3]. At first sight, *also indeed*, as claimed by Barnett and Loudon, "it is difficult to see how any component of our derivation could seriously be open to question". However on second thoughts, one may find that this approach itself has implicitly assumed that, once the Abraham momentum and energy satisfy the global momentum-energy conservation law in one inertial frame of reference, then they will do in all inertial frames. Obviously, there is no basis to support such an implicit assumption in the Einstein-box thought experiment.